\documentclass[11pt,twoside]{article}

\usepackage{asp2006}
\usepackage{epsf}
\usepackage{psfig}
\usepackage{lscape}

\begin{document}

\title{Mechanisms of the physical connection between the radio- and high-energy emissions of pulsars}
\author{S. A. Petrova}
\affil{Institute of Radio Astronomy, Chervonopraporna Str., 4, Kharkov 61002, Ukraine}    

\begin{abstract} 
The high-energy emission mechanisms based on the radio photon
reprocessing by the ultrarelativistic plasma particles in the open field
line tube of a pulsar are considered. The particles are believed to acquire relativistic gyration energies as a result of resonant absorption of pulsar radio emission. The spontaneous synchrotron re-emission of these particles falls into the optical and soft X-ray ranges and can at least partially account for the pulsar non-thermal high-energy emission. Besides that, the radio photons, which are still below the resonance, can be deposited into the high-energy range by means of the scattering off the gyrating particles. This process can also markedly
contribute to the pulsar high-energy emission and underlie the potentially observable features of the radio -- high-energy connection. Based
on the theory developed, we interpret the manifestations of the
radio -- high-energy connection already observed in the Crab and Vela
pulsars. Furthermore, it is shown that generally the most prominent connection is expected at the lowest radio frequencies, beyond the
low-frequency turnover of a pulsar.
\end{abstract}

\section{Introduction}
The radio and high-energy emissions of pulsars have substantially distinct energetics and spectra, and they are undoubtedly generated by distinct mechanisms. The high-energy emission is believed to originate in the regions of particle acceleration, in the polar, outer or slot gap, while the radio emission region is usually placed deep in the tube of open magnetic lines, but well above the polar gap. In this picture, the emission mechanisms are independent. However, the simultaneous observations in radio and at high energies have already revealed joint fluctuations in these ranges, testifying to the physical connection between the radio and high-energy emission of pulsars.

In the Crab pulsar, the optical pulses coincident with giant radio pulses are 3\% brighter than the average pulse \citep{p1-s03,p1-o08}. The Vela pulsar does not show classical giant pulses, but it is well known that its radio profile changes with radio intensity \citep{p1-kd83}. As the intensity increases, the role of the leading component (the so-called precursor), increases as well, while the rest of the profile (the main pulse) vanishes. Surprisingly, the soft X-ray profile of the Vela pulsar integrated over the range 2-16 keV also changes with radio intensity \citep{p1-l07}. The changes touch the main peak of the profile and the so-called trough in the leading region. Note that at softer energies this trough presents a separate component which passes through its maximum at a few tenths keV \citep{p1-h02}.

The observed manifestations of the radio -- high-energy connection imply that
{\itshape i)} this connection is restricted to soft enough energies, {\itshape ii)}
the high-energy fluctuations correlated with radio emission are very weak as compared to the radio pulse changes, and {\itshape iii)} the high-energy fluctuations touch not only the total flux, but also the profile shape. Therefore this connection can hardly be directly attributed to the global electrodynamic changes in the magnetosphere. It can rather be interpreted in terms of propagation effects.

\section{Physics of radio photon reprocessing in pulsar magnetosphere}

As the radio beam propagates through the secondary plasma in the open field line tube, it can be reprocessed to high energies. In the radio emission region, the magnetic field is so strong that any perpendicular momentum is almost immediately lost via synchrotron re-emission, and the radio frequency in the particle's rest frame is much less than the electron gyrofrequency, $\omega^\prime\ll\omega_H\equiv eB/mc$. As the magnetic field strength decreases with altitude, $B\propto r^{-3}$, in the outer magnetosphere the radio waves pass through the cyclotron resonance, $\omega^\prime =\omega_H$. In the presence of resonant radiation, the particles can absorb and emit photons. In total, the incident radio emission is partially absorbed, while the particles increase their transverse momenta. Very soon their gyration becomes relativistic, and their transverse and total momenta continue growing \citep{p1-lp98,p1-p02}. The magnetic field in the resonance region is much weaker than that in the radio emission region, and the synchrotron re-emission no longer prevents the momentum growth. 

The synchrotron radiation of the particles with the evolved momenta falls into the optical and soft X-ray ranges, and it can be detectable. This was suggested as a mechanism of the pulsar high-energy emission \citep*{p1-lp98,p1-g01,p1-p03}. Later on this model was elaborated in \citet*{p1-h05,p1-h08} by including the effect of the accelerating electric field on the particle momenta. In the framework of this model, the physical connection between the radio and high-energy emission of pulsars is natural. The excess of optical emission during giant radio pulses in the Crab pulsar can be understood as follows. As a giant radio pulse comes into the resonance region, the momentum evolution of the particles becomes more pronounced, and their synchrotron emission is stronger.

To interpret the correlation observed in the Vela pulsar it is necessary to include one more effect of the radio photon reprocessing to high energies \citep{p1-p09a}. Pulsar radio emission is essentially broadband, and therefore the resonance region is quite extended. Over most part of this region there is a significant amount of the under-resonance photons, with frequencies $\omega^\prime\ll\omega_H$. The particles acquire relativistic gyration at the very bottom of the resonance region, and further on the under-resonance photons can be scattered off the relativistically gyrating particles. This process differs essentially from the common magnetized scattering by straightly moving particles \citep{p1-p08}. The under-resonance photons are chiefly scattered to high harmonics of the particle gyrofrequency, $\omega_{\rm sc}^\prime =\omega^\prime +s\omega_H$ with $s\sim\gamma_0^3$ (where $\gamma_0$ is the Lorentz-factor of the particle gyration), and the total scattering cross-section is much larger. Figure 1 shows the spectral distribution of the scattered power as compared to the synchrotron spectrum of the same particle. One can see that the peak of the scattered radiation is markedly shifted beyond the synchrotron maximum. Although the total power scattered is always less than the synchrotron power, in the region beyond the synchrotron maximum the contribution of the scattered radiation can be substantial. 

\begin{figure}
\plotone{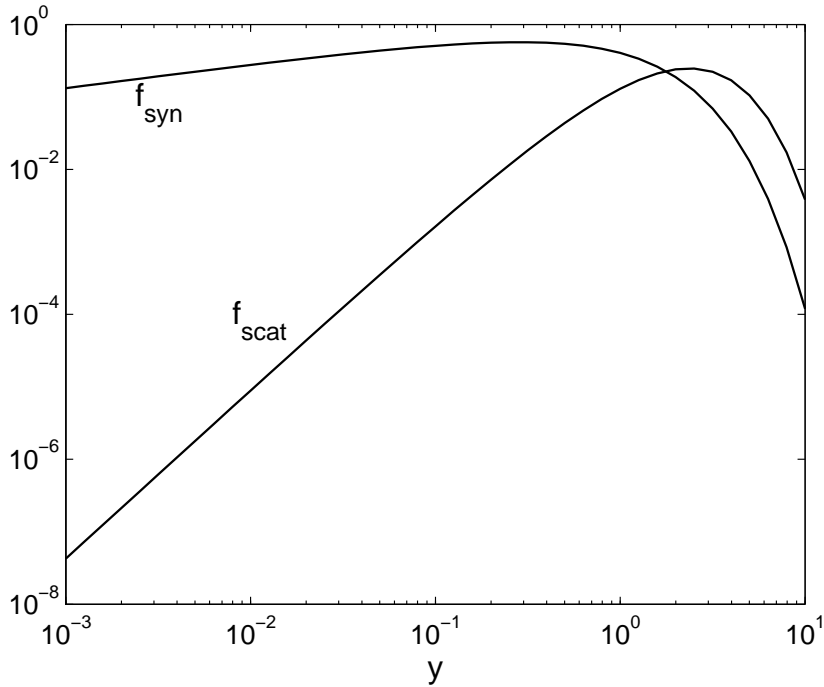}
\caption{Spectral distributions of the scattered and synchrotron powers of the particle, $y\equiv (2/3)s\gamma_0^{-3}$.}
\end{figure}

It is important to note that the scattered power is a very strong function of the particle gyration energy, $L_{\rm sc}\propto\gamma_0^6$, and the quantity $\gamma_0$ is determined by the radio intensity coming into the resonance region and causing the particle momentum evolution. Thus, it is the scattered power that is tightly connected with the radio emission characteristics, and this connection should be most prominent beyond the synchrotron maximum. Recall that in the Vela pulsar the trough component is indeed noticeably connected with the radio emission beyond its spectral maximum. The numerical estimates as applied to this pulsar give reasonable results. In the observer frame, the peaks of the synchrotron and scattered radiation, $\hbar\omega_{\rm syn}=0.2$ keV and $\hbar\omega_{\rm sc}=1.5$ keV, well agree with the spectral maximum of the trough component and the range of its pronounced correlation with radio, while the total synchrotron power, $L_{\rm syn}=10^{31}$ erg\,s$^{-1}$, is compatible with the observed luminosity of this component.

The radio beam is also subject to induced scattering off the gyrating particles. In contrast to the spontaneous scattering considered above, the induced scattering is most efficient between the states below the resonance, $\omega_{\rm sc}^\prime=\omega^\prime$, and can modify the radio profile shape \citep[for more details see][and references therein]{p1-p09b}. Here it should only be mentioned that the efficiency of induced scattering is also a function of the particle gyration energy $\gamma_0$. Hence, the variations of $\gamma_0$ should lead to joint fluctuations in the radio and high-energy emission, and the radio -- high-energy connection in the Vela pulsar can be understood as the interplay between the processes of spontaneous and induced scattering by the spiraling particles.

\section{Expected features of the radio -- high-energy connection at low radio frequencies} 

In our theory the physical connection of the radio and high-energy emission is typical of pulsars. Note that the resonant absorption and the spontaneous scattering by the gyrating particles both become much more efficient at low radio frequencies. Probably, these processes determine the low-frequency turnover in pulsar spectrum, which is observed in the population of old pulsars. If so, the original radio luminosity is much larger than the observed one, and the radio luminosity lacking beyond the turnover should appear at high energies. Obviously, the correlation between the radio emission beyond the turnover and the high-energy emission should be most pronounced. Note that in several old pulsars there is indeed the unexpected excess of the non-thermal high-energy emission \citep[e.g.][]{p1-beck}.

The propagation origin of the turnover should imply peculiar radio intensity statistics at lower frequencies. Even small variations in the plasma flow may cause substantial change of the efficiency of reprocessing and affect the outgoing radio intensity drastically. Figure 2 shows the results of numerical simulations of the final single-pulse intensities at two radio frequencies as well as the total radio spectrum. The original power-law spectrum is also given here. One can see that the distribution of the single-pulse intensities beyond the turnover is essentially asymmetric, in contrast to that at high radio frequencies. The radiation beyond the turnover is dominated by several strong pulses, while the overwhelming majority of the pulses are well below the average. This trend is compatible with the observed statistics of the low-frequency giant pulses \citep[e.g.][]{p1-kuzmin} and the anomalously strong pulses \citep{p1-ulyanov} found beyond the turnover.

\begin{figure}[!ht]
\plotone{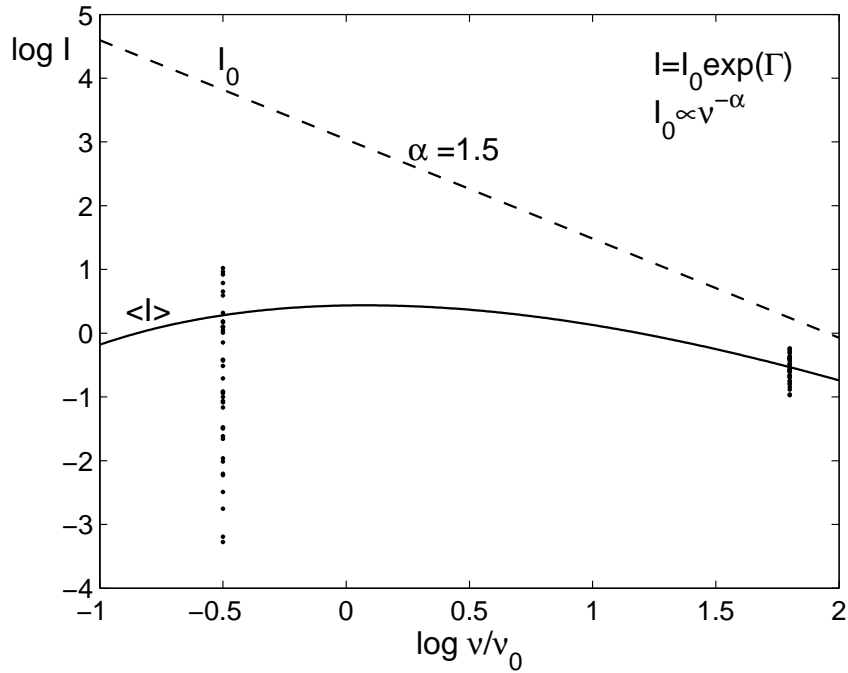}
\caption{Simulated radio spectrum of a pulsar with account for the radio photon reprocessing to high energies (solid line), the original spectrum (dashed line), and the single-pulse intensities at two radio frequencies.}
\end{figure}

\section{Conclusions}
Resonant absorption of pulsar radio emission in the tube of open magnetic lines leads to relativistic gyration of the absorbing particles and makes possible the radio photon reprocessing to high energies. The scattering of the under-resonance radio photons by the gyrating particles and their synchrotron re-emission can both have observable consequences in a number of pulsars and imply joint fluctuations of the radio and high-energy emission. The radio -- high-energy correlation is expected to be most pronounced at low radio frequencies, below the low-frequency turnover in pulsar spectra.

\acknowledgements I am grateful to the LFRU organizers for the financial assistance which has made possible my participation in the Conference. The work is partially supported by INTAS Grant No 03-5727.

\end{document}